\documentclass[nospthms,smallextended,final]{svjour2}

\usepackage{color}
\usepackage{graphicx}

\usepackage{amsmath}
\usepackage{amssymb}

\newcommand{\Rey}{\mathrm{Re}}
\newcommand{\la}{\langle}
\newcommand{\ra}{\rangle}
\newcommand{\Ei}{\langle E_i \rangle}
\newcommand{\Eip}{\langle E_{i+1} \rangle}
\newcommand{\Eim}{\langle E_{i-1} \rangle}
\newcommand{\ve}{\varepsilon}
\newcommand{\mE}{\mathcal{E}}
\newcommand{\Eturb}{E_1^{\mathrm{turb}}}

\begin{document}

\title{A cascade model for the discontinuous absorbing phase transition between turbulent and laminar flows}
\date{\today}

\author{Eric Bertin \and Alex Andrix \and Ga\"el Le Godais}
\institute{Eric Bertin \at Universit\'e Grenoble Alpes, CNRS, LIPhy, 38000 Grenoble, France
\at Institut Rh\^one-Alpin des Syst\`emes Complexes, IXXI, F-69342 Lyon, France
\\ \email{eric.bertin@univ-grenoble-alpes.fr}
\and Alex Andrix \at Institut Rh\^one-Alpin des Syst\`emes Complexes, IXXI, F-69342 Lyon, France
\\ 
\and Ga\"el Le Godais \at Institut Rh\^one-Alpin des Syst\`emes Complexes, IXXI, F-69342 Lyon, France
\\ 
}
\maketitle

\begin{abstract}
We introduce a minimal model of energy transfer through scales to describe, at a qualitative level, the subcritical transition between laminar and turbulent flows,
viewed in a statistical physics framework as a discontinuous absorbing phase transition.
The main control parameter of the model is a Reynolds number that compares energy transfer to viscous dissipation on a large length scale.
In spite of its simplicity, the model qualitatively reproduces a number of salient features of the subcritical laminar-turbulent transition, including the existence of an absorbing laminar state,
the discontinuous onset of a metastable fluctuating turbulent state above a threshold Reynolds number, and a faster-than-exponential increase of the turbulence lifetime when increasing the Reynolds number.
The behavior of the model is also consistent, at high Reynolds number, with the Kolmogorov K41 phenomenology of fully developed turbulence.
\end{abstract}

\section{Introduction}

The beautiful natural scenery of a waterfall is an inspiring experience for both the artist and the scientist. In waterfalls where water rapidly flows down over the ground --instead of freely falling into the air--, one clearly sees a change of behavior between the quiet laminar flow of water upstream from the cascade, and the swirly and fluctuating turbulent flow in the cascade, that then progressively recovers its laminar state downstream from the waterfall.
Such a natural phenomenon is an example of the transition from laminar to turbulent flow, that has been extensively studied and characterized in controled laboratory experiments \cite{Tillmark92,Daviaud92,Prigent:2002fqa,Avila:2011up,Samanta:2011va}.
In this transitional regime, the turbulent flow significantly differs from fully turbulent flows obtained under strong external drive, where fluctuating structures span a broad range of length scales, and exhibit scale invariance \cite{Frisch}.
Transitional turbulence is rather a form of spatiotemporal chaos, involving a comparably narrow range of length scales \cite{Daviaud92}.
The control parameter quantifying the intensity of the drive, and allowing for a classification of the different regimes (e.g., laminar flow, transitional turbulence or fully developed turbulence), is the Reynolds number $\Rey$, which basically compares inertial energy transfer to viscous damping at large length scale. A fluid flow is typically laminar at low Reynolds number (where viscous effects dominate), while it is turbulent at high Reynolds number (where inertial effects dominate) \cite{Frisch}.

Early characterizations of the transition to turbulence were based on dynamical systems approaches, and started by a linear stability analysis of the laminar state, where no fluctuations are present. For some flow geometries, like the Taylor-Couette cylindrical flow, the laminar state becomes linearly unstable above a critical Reynolds number $\Rey_c$, where a supercritical bifurcation takes place \cite{Joseph:1976twa,Grossmann:2000zz,Dauchot:1997wta}. For $\Rey>\Rey_c$, small fluctuations around the laminar state grow exponentially and lead to the formation of dissipative structures like vortices, which may themselves become unstable at higher Reynolds number through a bifurcation cascade, eventually leading to a spatiotemporally chaotic state \cite{DiPrima1985}.
For other geometries like the sheared plane Couette flow between two plates moving in opposite directions \cite{Daviaud92}, or the Poiseuille pipe flow driven by a pressure gradient \cite{Avila:2011up}, the laminar flow is linearly stable up to very high values of Reynolds number \cite{Romanov73,Salwen1980}. In this case, the turbulent state appears through a subcritical bifurcation and can only be reached through finite amplitude perturbations around the laminar state \cite{Dauchot:1995tha,Avila2023}. The minimal perturbation amplitude beyond which the turbulent state is triggered typically decays as an inverse power law of the Reynolds number \cite{Dubrulle1991,Kreiss1994}, so that at high Reynolds number experimental noise is enough, in practice, to destabilize the laminar state.
In this transitional regime, turbulent spots coexists with laminar regions, leading to a rich spatiotemporal dynamics \cite{Carlson82,Tillmark92,Daviaud92,Dauchot:1995tha}.

In large aspect ratio experiments \cite{Prigent:2002fqa,Prigent:2003uma,borrero2010,Avila:2011up,Samanta:2011va} or numerical simulations \cite{Barkley2005,Tuckerman:2008gw,Duguet:2009bd,Schneider:2010id,Duguet:2010dv,Manneville:2010hd,Philip:2011ja} in the plane Couette geometry, the flow organizes into turbulent bands with a well-defined length scale, that has been shown to result from a linear instability (in a statistical sense) of the homogeneous turbulent state, that occurs when decreasing the Reynolds number \cite{Dauchot2022}.
The coexistence of laminar and turbulent regions, as well as the need for finite amplitude perturbations to reach the turbulent state from the laminar one, suggests a qualitative analogy with discontinuous phase transitions (e.g., the equilibrium liquid-gas phase transition). However, here the transition occurs far from equilibrium, and rather resembles absorbing phase transitions (albeit of a discontinuous type), with the laminar state playing the role of the absorbing state and the turbulent state corresponding to the active one \cite{Pomeau86,sipos2011directed,Hof2022}.
The turbulent state has a finite lifetime which increases faster than exponentially with the Reynolds number, close to the laminar-turbulent transition \cite{Hof:2006fk,Goldenfeld2010}.
The analogy with absorbing phase transitions was already pointed out in the seminal work of Pomeau \cite{Pomeau86}, who suggested that the dynamics of turbulent and laminar regions may be described in terms of directed percolation (the prominent class of absorbing phase transition), a conjecture recently confirmed experimentally \cite{Hof2022}.

Apart from a model inspired by an unexpected analogy with the physics of glasses \cite{Bertin2012,Bertin2014}, models of the transition to turbulence have so far focused on the description of the complex spatio-temporal dynamics, and often take the form of coupled (partial) differential equations (the Navier-Stokes equations, or projections thereof), whose behavior beyond the laminar state can only be studied numerically \cite{Eckhardt:2008gd,MannevilleEPJ07,Gibson:2008ec,Barkley2011}. This is in contrast with fully developed turbulence, where toy models amenable to analytical treatments have been proposed and analyzed.
In particular, the turbulent cascade phenomenology of fully developed turbulence can be qualitatively described using shell models \cite{Gledzer1973,Yamada1988,LVov1998,Ditlevsen}, or even, at a metaphoric level, using simpler stochastic models of energy transfer through scales \cite{Bertin08}.

Getting inspiration from both shell models for turbulence \cite{Ditlevsen} and stochastic models describing absorbing phase transitions \cite{Hinrichsen2000}, we propose and analyze a stochastic energy transfer model which qualitatively captures a number of key properties of the subcritical transition to turbulence:
(i) the existence of an absorbing laminar state for all Reynolds number;
(ii) the existence of a fluctuating turbulent state above a characteristic Reynolds number;
(iii) a faster-than-exponential increase of the turbulent lifetime with Reynolds number;
(iv) a perturbation amplitude to reach the turbulent state from the laminate state which decays as a power law of the Reynolds number;
(v) a behavior consistent with the fully developed turbulence phenomenology in the limit of high Reynolds number.
We show that such a model can be derived using minimal phenomenological assumptions, and we characterize its main properties.

\section{Schematic shell model for the laminar-turbulent transition}

\subsection{Generic form of the stochastic shell model}

It is customary to represent the velocity field of the fluid in terms of spatial Fourier modes, labeled by a wavevector $\mathbf{k}$.
In the spirit of shell models for fully-developed turbulence, we consider that the wavevector space can be split into $N+1$ different shells,
associated with a characteristic wavenumber $k_i$ ($i=0,\dots,N$) \cite{Ditlevsen}. We assume, again as in standard shell models,
that the wavenumbers $k_i$ are exponentially spaced, i.e., $k_i=k_0 \lambda^i$, with $\lambda>1$ a characteristic scale ratio.
The energy $E_i(t)$ of shell $i$ aggregates the energies of all Fourier modes whose wavevector $\mathbf{k}$ is within shell $i$
($1< |\mathbf{k}|/k_i <\lambda$) \cite{Ditlevsen}. We assume throughout the paper that $E_i(t)\ge 0$.

As a minimal model, we assume that the drive maintains constant the energy $E_0$ of the shell $i=0$, corresponding to the largest length scale in the system, at which energy is injected (e.g., by applying an external shear). The constant energy $E_0$ thus becomes a control parameter of the model.
For $i=1,\dots,N-1$, the energy $E_i(t)$ is assumed to satisfy the following dynamics:
\begin{equation} \label{eq:dEi:dt}
    \frac{dE_i}{dt} = - \nu k_i^2 E_i + T_i(E_{i-1},E_i) - T_{i+1}(E_i,E_{i+1}) +  g_i(E_i)\, \xi_i(t).
\end{equation}
The term $-\nu k_i^2 E_i$ in the rhs of Eq.~(\ref{eq:dEi:dt}) accounts for viscous dissipation, $\nu$ being the viscosity.
The term $T_i(E_{i-1},E_i)$ models in a simplified way the energy transfer rate from shell $i-1$ to shell $i$, due to nonlinear (inertial) terms in the Navier-Stokes equation. For the sake of simplicity, energy transfer is assumed to be fully biased from shell $i-1$ to shell $i$, with no back flux 
from shell $i$ to shell $i-1$. As a boundary condition on the shell $i=N$, we assume for simplicity a constant energy $E_N$, taking a zero or small positive value.
The last term in the rhs of Eq.~(\ref{eq:dEi:dt}) is a multiplicative Langevin noise (in the It\={o} interpretation) inducing fluctuations in the turbulent phase,
where $\xi_i(t)$ is a Gaussian white noise satisfying $\langle \xi_i(t) \rangle =0$ and $\langle \xi_i(t) \xi_j(t') \rangle = \, \delta_{ij} \delta(t-t')$.
The noise models the effect of fast chaotic degrees of freedom that are not explicitly retained in the description.
Note that a Langevin noise was also used in a Ginzburg-Landau description of the turbulent state in \cite{Prigent:2002fqa,Prigent:2003uma} to model turbulent fluctuations, but the noise was additive instead of multiplicative, as the model considered did not aim at capturing both the laminar and turbulent states.
The use of a multiplicative Langevin noise is a standard practice in the description of absorbing phase transitions, to ensure that noise vanishes in the absorbing state \cite{Hinrichsen2000}.
We thus require that $g_i(0)=0$. The specific form chosen for the function $g_i(E_i)$ is discussed below.

\subsection{Dimensionally-consistent energy transfer terms}

To model both the energy transfer term (or energy flux) $T_i(E_{i-1},E_i)$ and the noise amplitude $g_i(E_i)$, we resort to dimensional analysis.
By construction, $T_i(E_{i-1},E_i)$ can only depend on the energies $E_{i-1}$, $E_i$, and on the wavenumber $k_i$.
The energies have dimension $[E_j]=L^2 T^{-2}$, while the energy flux has dimension $[T_i(E_{i-1},E_i)]=L^2 T^{-3}$,
where $L$ and $T$ indicate the dimensions of length and time respectively \cite{Ditlevsen}.
Hence a dimensionally-consistent expression of the energy flux necessarily takes the form
\begin{equation} \label{eq:transfer:DA}
    T_i(E_{i-1},E_i) = k_i E_i^{3/2} f_i\left(\frac{E_i}{E_{i-1}}\right),
\end{equation}
where $f_i(x)$ is at this stage an arbitrary positive dimensionless function of the dimensionless argument $x$.
In the following, we make the further simplification that $f_i(x) \equiv f(x)$ is the same for all shells.
A brief discussion of the qualitative analogy with energy transfer terms in the Navier-Stokes equation is given in Appendix~\ref{sec:app:A}.

\subsection{Dimensionally-consistent multiplicative noise term}

Turning to the noise amplitude $g_i(E_i)$, we assume that it depends only on $k_i$ and $E_i$. Dimensional arguments thus lead us to postulate the following power-law form
\begin{equation} \label{eq:scaling:giEi}
g_i(E_i) = \eta k_i^{\delta} E_i^{\gamma}\,,
\end{equation}
where $\eta$ is a dimensionless constant, and $\delta$, $\gamma$ are exponents to be determined.
As the noise correlation has dimension $[\langle \xi_i(t) \xi_i(t') \rangle] = [\delta(t-t')] = T^{-1}$, the noise thus has dimension $[\xi_i(t)]=T^{-1/2}$.
Since
\begin{equation}
[g(E_i)\xi_i(t)] = \left[ \frac{dE}{dt} \right] = L^2 T^{-3},
\end{equation}
it follows that $[g(E_i)]=L^2 T^{-5/2}$. Recalling that $[E_i]=L^2 T^{-2}$, we get from Eq.~(\ref{eq:scaling:giEi}), by identification, that
$\gamma=\frac{5}{4}$ and $\delta=\frac{1}{2}$.

\subsection{Dimensionless dynamical equations}

The transition between laminar and turbulent states is controlled by the Reynolds number, that quantifies the intensity of the external drive by comparing the order of magnitude of the energy transfer by inertial terms to the viscous damping on a large length scale. In the present model, we thus define the Reynolds number as
\begin{equation} \label{eq:def:Rey}
    \Rey = \frac{k_1 E_0^{3/2}}{\nu k_1^2 E_0} = \frac{\sqrt{E_0}}{\nu k_1}\,,
\end{equation}
by comparing the energy transfer to mode $i=1$, roughly estimated as $k_1 E_0^{3/2}$ (by assuming that $E_1$ is of the order of $E_0$ and that $f(1)\approx 1$), to the typical scale of the viscous damping, estimated as $\nu k_1^2 E_0$.
Defining the dimensionless shell energies $\tilde{E}_i=E_i/E_0$, the dimensionless time $\tilde{t}=t k_1 \sqrt{E_0}$, and the dimensionless wavenumber $\tilde{k}_i=k_i/k_1$, 
Eq.~(\ref{eq:dEi:dt}) turns into,
also using Eqs.~(\ref{eq:transfer:DA}) and (\ref{eq:scaling:giEi}):
\begin{align} \label{eq:dEi:dt:adim}
    \frac{d\tilde{E}_i}{dt} &= - \frac{1}{\Rey} \tilde{k}_i^2 \tilde{E}_i + \tilde{k}_i \tilde{E}_i^{3/2} f\left(\frac{\tilde{E}_i}{\tilde{E}_{i-1}}\right)\\ \nonumber
    & \qquad \qquad \qquad - \tilde{k}_{i+1} \tilde{E}_{i+1}^{3/2} f\left(\frac{\tilde{E}_{i+1}}{\tilde{E}_i}\right) + \eta \tilde{k}_i^{1/2} \tilde{E}_i^{5/4} \, \tilde{\xi}_i(\tilde{t}),
\end{align}
for $i=1,\dots,N-1$, and with $\tilde{E}_0=1$.
From now on, we work with dimensionless equations, and we drop the tilde on dimensionless quantities to lighten notations.


\section{Deterministic dynamics under a mean-field approximation}

\subsection{Local mean-field approximation}

We start by considering a local mean-field approximation of the dynamics, by taking an ensemble average of Eq.~(\ref{eq:dEi:dt}), which removes the noise term,
and by approximating averages of the transfer terms, $\la T_i(E_{i-1},E_i)\ra$, by $T_i(\la E_{i-1}\ra,\la E_i\ra)$. We then get
\begin{equation} \label{eq:dEi:dt:MF}
    \frac{d\Ei}{dt} = - \frac{1}{\Rey} k_i^2 \Ei + k_i \Ei^{3/2} f\left(\frac{\Ei}{\Eim}\right) - k_{i+1} \Eip^{3/2} f\left(\frac{\Eip}{\Eim}\right),
\end{equation}
recalling that in dimensionless form, $k_i=\lambda^{i-1}$.
The approximate description given in Eq.~(\ref{eq:dEi:dt:MF}) has the advantage of being more easily amenable to analytical treatments, in particular to study laminar and turbulent states which both appear as fixed points of Eq.~(\ref{eq:dEi:dt:MF}).

\subsection{Consistency with K41 theory of fully-developed turbulence}

Before studying the transitional regime between laminar and turbulent states, we note that the mean-field evolution equation (\ref{eq:dEi:dt:MF}) is compatible with the Kolmogorov K41 theory of
fully developed turbulence \cite{Frisch}. For very high Reynolds number, the viscous dissipation term $-k_i^2 \Ei/\Rey$ can be neglected over the inertial range (i.e., as long as
the wavenumber $k_i$ remains smaller than the so-called dissipative scale \cite{Frisch}) with respect to the energy transfer terms.
In the inertial range, the stationary state of Eq.~(\ref{eq:dEi:dt:MF}) corresponds to a constant energy flux $\ve$,
\begin{equation} \label{eq:dEi:dt:MF:K41}
    k_i \Ei^{3/2} f\left(\frac{\Ei}{\Eim}\right) = \ve
\end{equation}
where $\ve$ is independent of $i$. It follows that
\begin{equation}
    \Ei = \zeta\, \ve^{2/3} k_i^{-2/3}
\end{equation}
with $\zeta=f(\lambda^{-2/3})^{-2/3}$, and $\lambda=k_i/k_{i-1}$. 
The energy $\Ei$ corresponds to the average energy integrated over a wavenumber shell of width $\Delta k_i = k_{i+1} - k_i=k_i (\lambda-1)$, taking into account that $k_i=k_0 \lambda^i$.
One thus finds for the energy spectral density,
\begin{equation}
    \frac{\Ei}{\Delta k_i} \propto \ve^{2/3} k_i^{-5/3},
\end{equation}
consistently with the Kolmogorov K41 theory.
This comes from the fact that we have taken into account the same type of dimensional argument as in K41 theory \cite{Frisch} when defining the transfer rates in Eq.~(\ref{eq:transfer:DA}).

\subsection{Choice of energy transfer terms from stability criteria}

We now determine the key properties of the function $f(x)$ appearing in the expression (\ref{eq:transfer:DA}) of the energy transfer terms,
based on the expected physical properties of the model. Let us assume for now that $\Ei=0$ for $i=2,\dots, N$.
Eq.~(\ref{eq:dEi:dt:MF}) specified to $i=1$ then boils down to
\begin{equation} \label{eq:dEi:dt:MF:i2}
    \frac{d\la E_1 \ra}{dt} = - \frac{1}{\Rey} \la E_1 \ra + \la E_1 \ra^{3/2} f(\la E_1 \ra)\,.
\end{equation}
Under the above assumptions, the state $\la E_1\ra =0$ corresponds to the laminar state, and it is indeed a fixed point of Eq.~(\ref{eq:dEi:dt:MF:i2}),
provided that the function $x^{3/2}f(x)$ goes to zero when $x\to 0$.
The linear stability of the laminar state depends on the asymptotic form of the function $f(x)$ for $x\to 0$. We assume that, in this limit,
$f(x) \sim a x^{-\alpha}$, with $a>0$ a constant. For small $x$, Eq.~(\ref{eq:dEi:dt:MF:i2}) then reads
\begin{equation} \label{eq:dEi:dt:MF:i2:lam}
    \frac{d\la E_1 \ra}{dt} = - \frac{1}{\Rey} \la E_1 \ra + a \la E_1 \ra^{3/2-\alpha}.
\end{equation}
For $\alpha > \frac{1}{2}$, the term proportional to $\la E_1 \ra^{3/2-\alpha}$ dominates over the linear term when $\la E_1 \ra \to 0$, and the laminar state $\la E_1 \ra=0$ is unstable for all values of Reynolds number $\Rey$.
This situation is not physically relevant, as in real fluid flows, the laminar state is always stable at low enough Reynolds number.
The case $\alpha = \frac{1}{2}$ corresponds to a supercritical bifurcation. In this case, Eq.~(\ref{eq:dEi:dt:MF:i2:lam}) boils down to
\begin{equation} \label{eq:dEi:dt:MF:i2:lam:supercrit}
    \frac{d\la E_1 \ra}{dt} = \left( a-\frac{1}{\Rey} \right) \la E_1 \ra,
\end{equation}
so that the laminar state $\la E_1 \ra=0$ is linearly stable for $\Rey<\Rey_c=a^{-1}$, and linearly unstable for $\Rey>\Rey_c$.
This supercritical bifurcation may be relevant, at a qualitative level, to describe specific flow geometries like the Taylor-Couette cylindrical flow.
However, in this case, the destabilization of the laminar state occurs through the onset of regular dissipative structures like vortices, that themselves become unstable at higher Reynolds number \cite{DiPrima1985}.
Our model is not designed to capture these regular dissipative structures, notably due to the aggregation of modes into shells, and to the presence of a stochastic noise mimicking the chaoticity of the turbulent state. Other modelling approaches, based on a more realistic deterministic description of the linear instability of the laminar state, would thus be better suited to describe the case of a subcritical bifurcation.

Finally, for $\alpha < \frac{1}{2}$, the leading term when $\la E_1 \ra \to 0$ is the linear term $-\la E_1 \ra /\Rey$, so that the laminar state $\la E_1 \ra=0$
is linearly stable for all values of the Reynolds number $\Rey$.
However, for high enough Reynolds number, a finite but small value of $\la E_1 \ra >0$ is enough to make the nonlinear transfer term approximated as $a \la E_1 \ra^{3/2-\alpha}$ dominate over the linear viscous damping term, in which case the energy $\la E_1 \ra$ grows with time, reaching a turbulent state to be characterized below.
This case corresponds to a subcritical bifurcation to turbulence, and is the main focus of this work. The subcritical transition scenario is of qualitative relevance to the plane Couette flow and the Poiseuille pipe flow, as well as to the Taylor-Couette flow in some regimes \cite{Prigent:2002fqa,Avila:2011up}.
The threshold amplitude $E_1^{\mathrm{th}}$ beyond which $\la E_1 \ra$ grows up to a turbulent state is obtained by balancing the two terms in the rhs of Eq.~(\ref{eq:dEi:dt:MF:i2:lam}), leading (again for $\alpha < \frac{1}{2}$)
to 
\begin{equation}
    E_1^{\mathrm{th}} = \left( \frac{1}{a\, \Rey} \right)^{2/(1-2\alpha)}.
\end{equation}
This result is consistent, at a qualitative level, with the fact that the threshold amplitude of perturbation is typically observed to decay with $\Rey$ as an inverse power law in subcritical transitions from laminar to turbulent flows \cite{Dubrulle1991,Kreiss1994}. 

Once $\la E_1 \ra > E_1^{\mathrm{th}}$, the energy $\la E_1 \ra$ grows, and this growth needs to be saturated to prevent a divergence of energy.
From Eq.~(\ref{eq:dEi:dt:MF:i2}), we see that the energy transfer term $\la E_1 \ra^{3/2} f(\la E_1 \ra)$ has to become sublinear for large $\la E_1 \ra$ so that viscous damping eventually dominates at large energy.
Assuming $f(x) \sim b x^{-\beta}$ ($b>0$) for $x\to +\infty$, a sublinear behavior of the transfer term corresponds to $\beta>\frac{1}{2}$.

As a simple parametrization of the function $f(x)$ that satisfies the asymptotic behaviors $f(x) \sim a x^{-\alpha}$ when $x\to 0$ and $f(x) \sim b x^{-\beta}$ for $x\to +\infty$,
we choose the form
\begin{equation} \label{eq:param:fx}
    f(x) = \frac{a}{x^{\alpha} \left(1+\frac{a}{b}x^{\beta-\alpha}\right)},
\end{equation}
assuming $\beta-\alpha>0$, a condition valid in the subcritical case where $\alpha<\frac{1}{2}$ and $\beta>\frac{1}{2}$.
The results below do not depend, at a qualitative level, on the details of the specific shape of $f(x)$ as long as this function smoothly interpolates between the two power-law asymptotic behaviors.

\subsection{Transition between laminar and turbulent states}
\label{sec:trans:turb:determ}

\begin{figure}[t]
    \centering
    \includegraphics[height=0.38\linewidth]{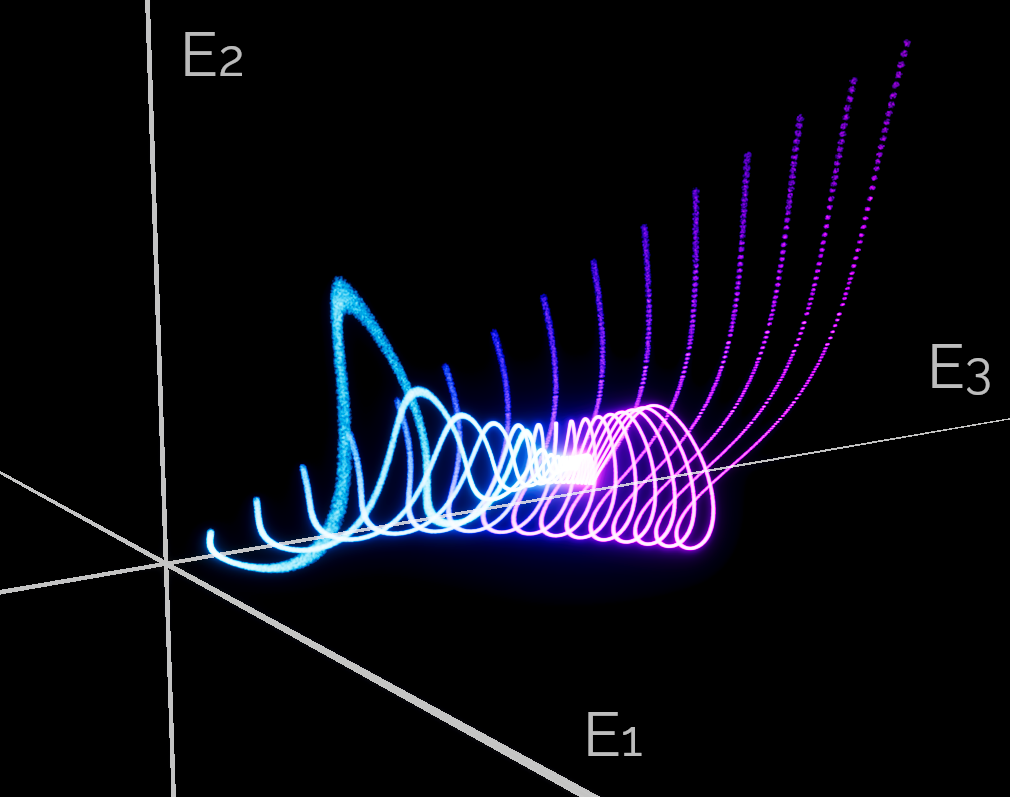}\hfill
    \includegraphics[height=0.38\linewidth]{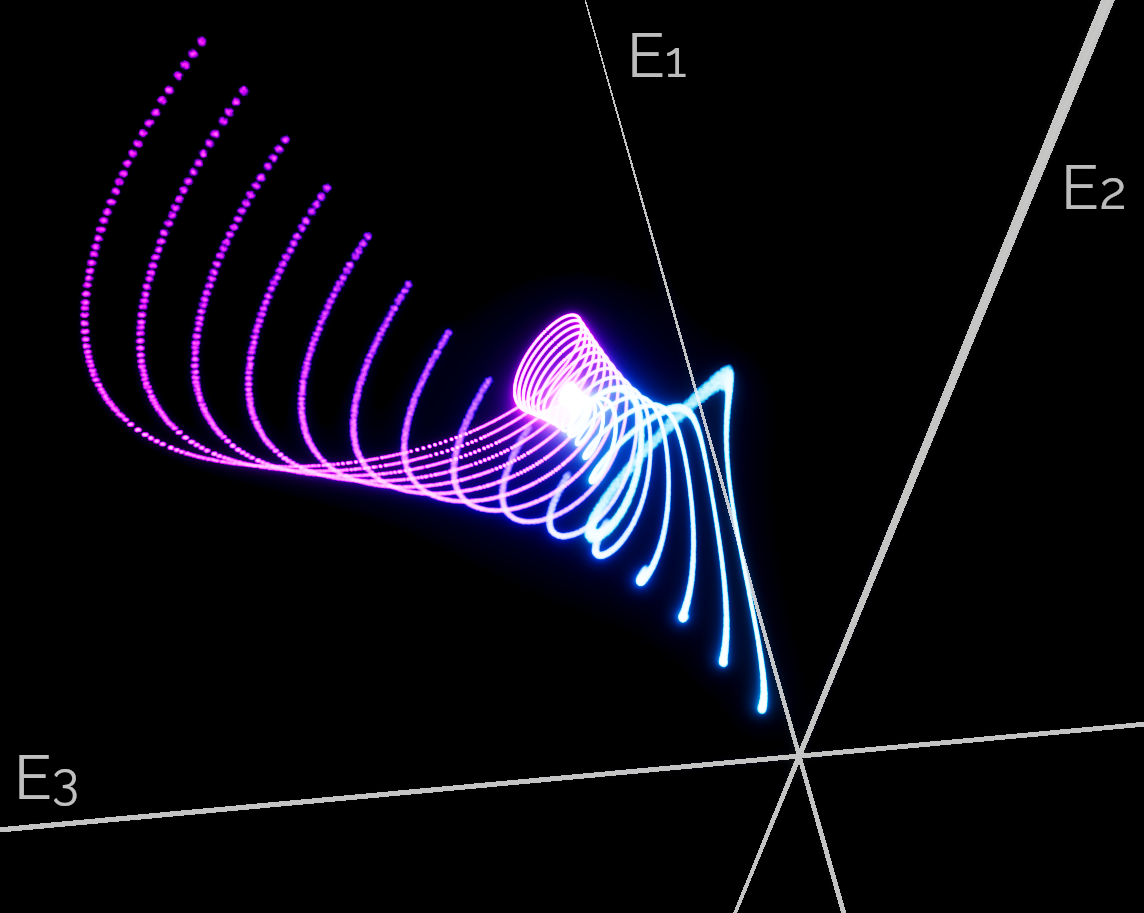}
    \caption{Illustrations of the trajectories converging to the turbulent fixed point in the ($E_1$, $E_2$, $E_3$)-space, starting from a set of distinct initial conditions, indicated by different colors. Both panels display the same trajectories, with different viewing angles. The brightness codes for the time along the trajectories. Parameters: $\Rey=13.3$, $\lambda=1.5$, $\eta=0$, $a=1$, $b=0.5$, $\alpha=0.25$, $\beta=1.25$, $N=4$. The energy $E_4$ is fixed to $E_4=10^{-3}$.}
    \label{fig1}
\end{figure}

According to Eq.~(\ref{eq:dEi:dt:MF:i2}), a stationary state with $\la E_1 \ra>0$ (to be interpreted as a turbulent state) satisfies
\begin{equation} \label{eq:dEi:dt:MF:i2:fixedpt}
    \la E_1 \ra^{1/2} f(\la E_1 \ra) = \frac{1}{\Rey}.
\end{equation}
With $f$ in the form (\ref{eq:param:fx}), the function
\begin{equation} \label{eq:def:Fx}
    F(\la E_1 \ra)=\la E_1 \ra^{1/2} f(\la E_1 \ra)
\end{equation}
has a maximum value reached for $\la E_1 \ra=E_1^*$, with
\begin{equation} \label{eq:def:E1star}
    E_1^*=\left[\frac{b\left(\frac{1}{2}-\alpha\right)}{a\left(\beta-\frac{1}{2}\right)}\right]^{1/(\beta-\alpha)}.
\end{equation}
The corresponding value $\Rey_t=F(E_1^*)^{-1}$ is the threshold Reynolds number for the existence of the turbulent state in the local mean-field approximation.
For $\Rey<\Rey_t$, Eq.~(\ref{eq:dEi:dt:MF:i2:fixedpt}) has no solution, meaning that the laminar state $\la E_1 \ra=0$ is the only stationary solution of Eq.~(\ref{eq:dEi:dt:MF:i2}).
In contrast, for $\Rey>\Rey_t$, Eq.~(\ref{eq:dEi:dt:MF:i2:fixedpt}) has two solutions, the threshold amplitude $E_1^{\mathrm{th}}$ and the turbulent state $\Eturb$ (with $E_1^{\mathrm{th}}<\Eturb$).
A linear stability analysis around these two fixed points shows that, as expected, $E_1^{\mathrm{th}}$ is an unstable fixed point whereas $\Eturb$ is a stable fixed point.
An illustration of the convergence of trajectories to the turbulent fixed under the deterministic dynamics is given in Fig.~\ref{fig1} (see also the corresponding movie provided as an Ancillary File).

The range of validity of the assumption that $\la E_i \ra=0$ for $i>1$ can be assessed recursively, in a self-consistent manner. Assuming that $\la E_i \ra=0$ for $i>2$, the dynamics of $\la E_2 \ra$
is similar to the dynamics of $\la E_1 \ra$ studied above, provided one redefines a mode-dependent Reynolds number $\Rey_2$ as
\begin{equation}
    \Rey_2 = \frac{\sqrt{\Eturb}}{\nu k_2}\,.
\end{equation}
This definition takes the same form as the original definition of the Reynolds number $\Rey$ given in Eq.~(\ref{eq:def:Rey}), upon replacement of the driving energy $E_0$ by the turbulent fixed point energy
$\la E_1\ra =\Eturb$, and of $k_1$ by $k_2$. As $\Eturb < E_0$ and $k_2>k_1$, one finds that $\Rey_2<\Rey$. More precisely, one has
\begin{equation} \label{eq:rel:Re:Re2}
    \Rey_2 = \frac{1}{\lambda} \left( \frac{\Eturb}{E_0} \right)^{1/2} \Rey.
\end{equation}
As long as $\Rey_2<\Rey_t$, the only fixed point of $\la E_2\ra$ is the laminar state, $\la E_2=0\ra$. Hence the assumption $\la E_i \ra=0$ for $i>1$ is consistent
in the range $\Rey_t < \Rey < \Rey_t'$, having defined
\begin{equation}
    \Rey_t' = \lambda \left( \frac{E_0}{\Eturb} \right)^{1/2} \Rey_t
\end{equation}
from the condition $\Rey_2=\Rey_t$, and using Eq.~(\ref{eq:rel:Re:Re2}). One easily checks that $\Rey_t'>\Rey_t$.


\section{Stochastic dynamics and turbulence lifetime}

In the previous section, we studied the deterministic dynamics of the model obtained under a mean-field assumption.
We now come back to the original Langevin dynamics defined in Eq.~(\ref{eq:dEi:dt:adim}).
Using Eqs.~(\ref{eq:transfer:DA}), (\ref{eq:scaling:giEi}) and (\ref{eq:param:fx}), the evolution equation for $E_i$ given in Eq.~(\ref{eq:dEi:dt:adim}) reads:

\begin{eqnarray} \label{eq:dEi:dt:full}
    \frac{dE_i}{dt} &=& - \frac{k_i^2}{\Rey} E_i \,+\, ab k_i E_i^{3/2} \left[ b\left(\frac{E_i}{E_{i-1}}\right)^{\alpha} + a\left(\frac{E_i}{E_{i-1}}\right)^{\beta} \right]^{-1}\\
    \nonumber
    && - \,ab k_{i+1} E_{i+1}^{3/2} \left[ b\left(\frac{E_{i+1}}{E_{i}}\right)^{\alpha} + a\left(\frac{E_{i+1}}{E_{i}}\right)^{\beta} \right]^{-1}
    \!\!+ \,\eta k_i^{1/2} E_i^{5/4}\, \xi_i(t).
\end{eqnarray}
We study below some properties of the turbulent state obtained for $\Rey>\Rey_t$.

\subsection{Representations of the turbulent state}

\begin{figure}[t]
    \centering
    \includegraphics[height=0.38\linewidth]{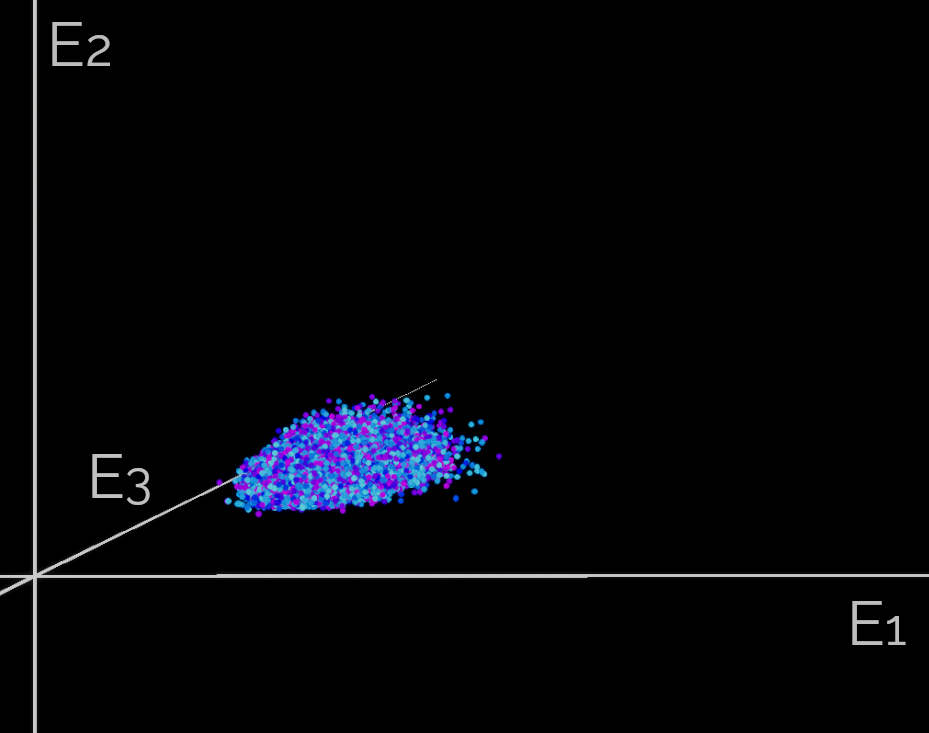}\hfill
    \includegraphics[height=0.38\linewidth]{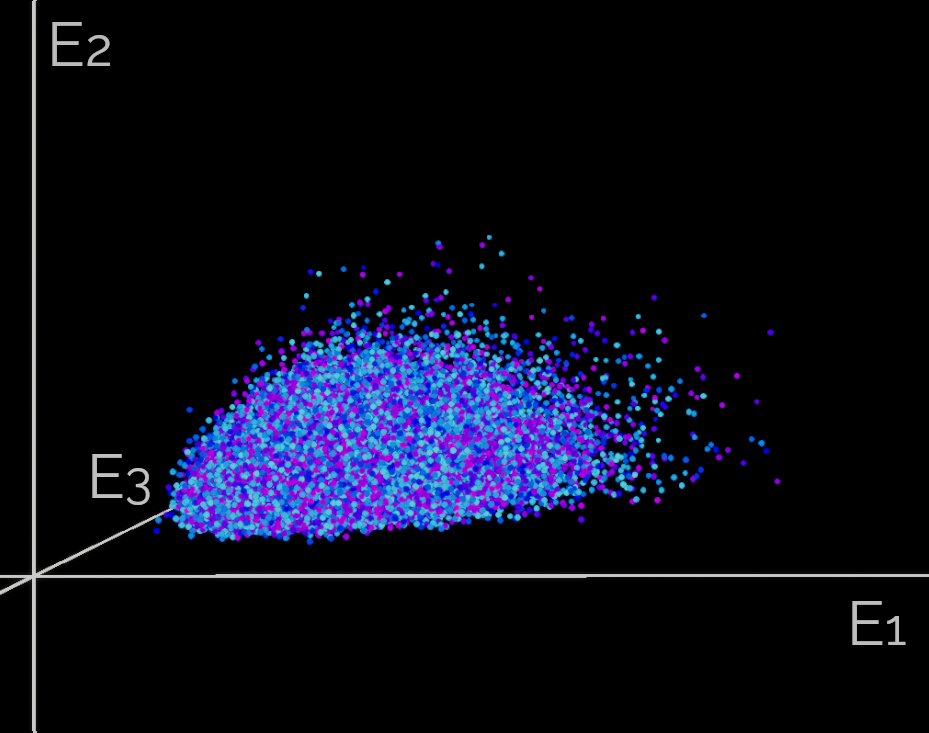}
    \caption{Illustration of the turbulent state of the stochastic dynamics in the ($E_1$, $E_2$, $E_3$)-space. Each particle represents a different realization of the stochastic dynamics. Same parameters as Fig.~\ref{fig1}, except $\eta=0.1$ for panel (a) and $\eta=0.2$ for panel (b).}
    \label{fig2}
\end{figure}

Under this stochastic dynamics, the turbulent state no longer appears as a fixed point, but rather as a fluctuating state, in qualitative agreement with experiments
\cite{Daviaud92,Bottin:1998ve,Prigent:2002fqa}.
These fluctuations can be visualized by considering many independent stochastic trajectories with statistically independent realizations of the noise $\xi_i(t)$.
In other words, one looks at many different independent copies of the system, which evolve in parallel.
In Fig.~\ref{fig2}, we plot the position of the resulting ensemble of copies in the $(E_1,E_2,E_3)$-space, using simulations with $N=4$,
and noise coefficient $\eta=0.1$ [Fig.~\ref{fig2}(a)] and $\eta=0.2$ [Fig.~\ref{fig2}(b)].
The resulting cloud of points provides a visualization of the fluctuations of the dynamics.

The turbulent state lasts only for a finite duration, after which the system eventually relaxes to the laminar state.
The turbulence lifetime strongly depends on the Reynolds number $\Rey$.
For $\Rey<\Rey_t$, the relaxation to the laminar state is very fast.
For $\Rey>\Rey_t$, the turbulent state emerges, and the system first relaxes to the turbulent (metastable) state before eventually relaxing to the laminar state, due to fluctuations.
To get some intuition of the relaxation process, we have represented the state of the system by an acoustic signal, by synthesizing an acoustic signal modulated by the modes $E_i$
(see Appendix~\ref{sec:app:B} for details on the synthesis procedure).
The resulting acoustic signal, determined for different values of Reynolds number (and two different noise amplitudes) is embedded in a movie, available as an Ancillary File. 
At $\Rey<\Rey_t$, the fast relaxation sounds like a fast decaying sound similar to some music instruments, while for $\Rey>\Rey_t$ the sound remains sustained as long as
the turbulent state has not relaxed to the laminar state. 

\subsection{Analytical determination of the turbulent lifetime close to the transition}

The mean-field determination of the turbulent state was done in Sec.~\ref{sec:trans:turb:determ} under the assumption that higher modes $\la E_i\ra$ ($i>1$) remain in the state $\la E_i\ra=0$, in which case the turbulent state can be characterized by the dynamics of $\la E_1 \ra$ only. This is true close to the transition Reynolds number $\Rey_t$,
for $\Rey_t < \Rey < \Rey_t'$. For higher Reynolds number, the turbulent state progressively involves additional modes, as illustrated in Fig.~\ref{fig2} for the stochastic dynamics
given by Eq.~(\ref{eq:dEi:dt:full}).
Here, we wish to determine analytically how the turbulent lifetime, which becomes finite in the case of the stochastic dynamics,
increases with Reynolds number close to the transition. 
We thus focus again on the dynamics of $E_1$, assuming $E_i=0$ for $i>1$. One expects the range of validity of this assumption to remain approximately the same as for the deterministic dynamics.

\begin{figure}[t]
    \centering
    \includegraphics[width=0.7\linewidth]{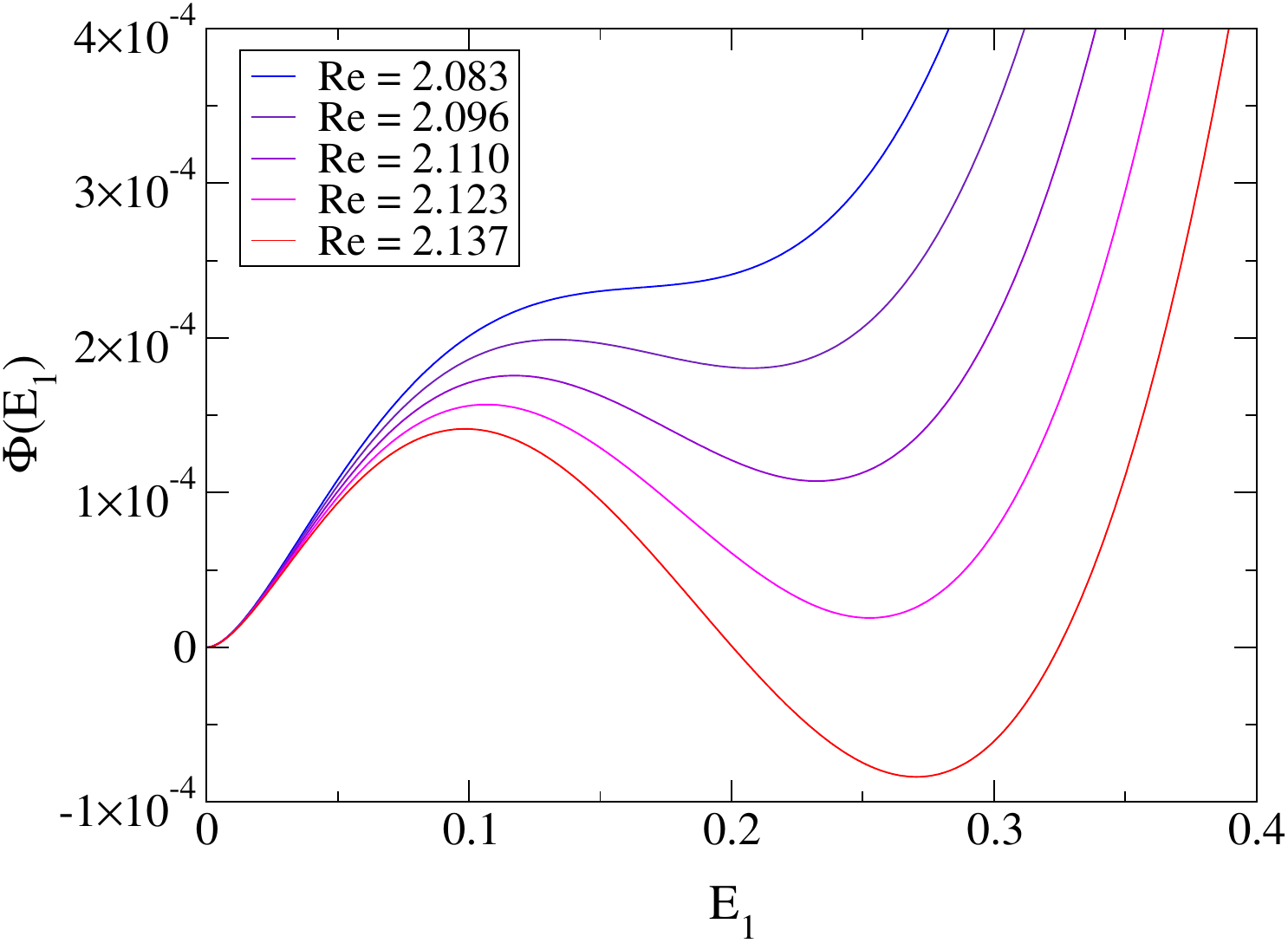}
    \caption{Effective potential $\Phi(E_1)$ for the stochastic dynamics of $E_1$, showing the emergence of the turbulent state as a local minimum of $\Phi(E_1)$ for
    $\Rey>\Rey_t=2.088$. Parameters: $a=1$, $b=0.5$, $\alpha=0.25$, $\beta=1.25$.}
    \label{fig:potential}
\end{figure}

Using Eq.~(\ref{eq:dEi:dt:full}) under the assumption that $E_i=0$ for $i>1$, the stochastic dynamics of $E_1$ can be written as
\begin{equation} \label{eq:dyn:stoch:E1:turb}
    \frac{dE_1}{dt} = -\Phi'(E_1) + \eta E_1^{5/4}\, \xi_1(t)
\end{equation}
(we recall that $k_1=1$ in dimensionless form), where the effective force $-\Phi'(E_1)$ derives from a potential
\begin{equation}
    \Phi(E_1) = \frac{E_1^2}{2\Rey} - \int_0^{E_1} dx \, \frac{a x^{3/2-\alpha}}{1+\frac{a}{b} x^{\beta-\alpha}}.
\end{equation}
The effective potential $\Phi(E_1)$ is plotted in Fig.~\ref{fig:potential} for different values of the Reynolds number, close to $\Rey_t$,
showing the onset of the turbulent state.
For $\Rey<\Rey_t$, the potential $\Phi(E_1)$ has a single minimum in $E_1=0$, corresponding to the laminar state.
In contrast, for $\Rey>\Rey_t$, a second minimum of the potential $\Phi(E_1)$ appears at the value $\Eturb$, corresponding to the turbulent state.
By increasing $\Rey$ further, the value $\Phi(\Eturb)$ of the potential in the turbulent state eventually becomes negative, i.e., $\Phi(\Eturb)<\Phi(0)=0$, where $\Phi(0)$ is the value
of the potential in the laminar state.
At equilibrium, the condition $\Phi(\Eturb)<\Phi(0)$ would lead one to conclude that the turbulent state then becomes the most stable state.
However, this is not the case here, due to the non-equilibrium character of the model, reflected in the presence of the multiplicative noise in Eq.~(\ref{eq:dyn:stoch:E1:turb}).
Indeed, since the amplitude of the noise vanishes for $E_1=0$, the laminar state $E_1=0$ is always a stable (i.e., absorbing) state, while $\Eturb$ remains a metastable state with a finite lifetime,
even when $\Phi(\Eturb)<\Phi(0)$.

To estimate the average lifetime of the laminar state close to the transition, we proceed as follows.
As a simplifying hypothesis, we assume that close to $\Rey_t$, the amplitude of the noise can be approximated by a constant amplitude
\begin{equation}
    \Gamma = \eta (E_1^*)^{5/4},
\end{equation}
assuming $E_1 \approx E_1^*$, where $E_1^*$ is defined in Eq.~(\ref{eq:def:E1star}).
Note that $E_1^*$ is approximately the mean value between the local maximum $E_1^{\mathrm{th}}$ of $\Phi(E_1)$ and the turbulent state $\Eturb$.
This approximation is not valid for the whole dynamics of $E_1$, but it is a reasonable assumption to describe the dynamics within the local potential well corresponding to the turbulent state,
since for $\Rey$ only slightly above $\Rey_t$, the extent of this local potential well is small.
Under this approximation, the computation of the average turbulent lifetime boils down to a standard Arrhenius escape problem over an energy barrier $\Delta \Phi$,
whose typical escape time scales as $\exp(\Delta \Phi/\Gamma)$, in the presence of a delta-correlated noise of amplitude $\Gamma$.

Defining $\varepsilon=\Rey-\Rey_t$, we expand $\Phi(E_1)$ around $E_1=E_1^*$, for small $\varepsilon$:
\begin{equation}
    \Phi(E_1^*+y) \approx \Phi(E_1^*) - \frac{E_1^*}{\Rey_t^2} \varepsilon y +  \frac{E_1^* \kappa}{3} y^3,
\end{equation}
with $|y|\ll E_1^*$, and where $\kappa=\frac{1}{2} |F''(E_1^*)|$; the function $F(E_1)$ is defined in Eq.~(\ref{eq:def:Fx}).
For $\varepsilon>0$, $\Phi_1(E_1^*+y)$ has a local minimum for $y=y_t>0$ (turbulent state) and a local maximum for $y=-y_t$, where
\begin{equation}
    y_t = \left( \frac{\varepsilon}{\kappa \Rey_t} \right)^{1/2}.
\end{equation}
One then obtains for the barrier $\Delta\Phi=\Phi(E_1^*-y_t)-\Phi(E_1^*+y_t)$:
\begin{equation}
    \Delta\Phi = \frac{4E_1^*}{3\kappa^{1/2} \Rey_t^3}\, \varepsilon^{3/2} \qquad (\varepsilon>0).
\end{equation}
As a result, the average turbulence lifetime $\tau$ behaves close to the laminar-turbulent transition as
\begin{equation}
    \tau \sim e^{\mu (\Rey-\Rey_t)^{3/2}},
\end{equation}
with $\mu=4E_1^*/(3\kappa^{1/2} \Rey_t^3)$.
Hence the turbulence lifetime increases faster than exponentially with the Reynolds number, in qualitative agreement with experimental results \cite{borrero2010}.

\section{Conclusion}

In this work, we have considered the subcritical transition to turbulence in the perspective of discontinuous absorbing phase transitions.
In this spirit, we have proposed and analyzed a stochastic energy transfer model which qualitatively captures a number of key properties of the subcritical transition to turbulence,
including the existence of an absorbing laminar state for all Reynolds number, the existence of a fluctuating turbulent state above a characteristic Reynolds number, and a faster-than-exponential increase of the turbulence lifetime with Reynolds number, close to the onset of the turbulent state.
The resulting model can be derived using minimal phenomenological assumptions, and is consistent with the Kolmogorov K41 phenomenology of fully developed turbulence in the limit of high Reynolds number.

However, due to its simple formulation in terms of aggregated modes over wavevector shells (in the spirit of shell models for isotropic fully developped turbulence), such a cascade model is not able to account for instance for the emergence of periodic turbulent bands reported in large-aspect-ratio experiments \cite{Prigent:2002fqa,Prigent:2003uma}
or numerical simulations \cite{Barkley2005,Tuckerman:2008gw} in the plane Couette and Taylor-Couette geometries.
The description of this more involved phenomenon would require to take into account the anisotropy of fluid flows exhibiting a subcritical transition to turbulence, which is certainly an interesting challenge for future work.

At any rate, the present simple model aims at tentatively bridging the gap between shell models for fully developed turbulence \cite{Ditlevsen} and simple models of absorbing phase transitions \cite{Hinrichsen2000}, since the relevance of the absorbing phase transition framework
to describe the subcritical laminar-turbulent transition has recently been confirmed experimentally \cite{Hof2022}. Along this line of thought, one might expect the laminar-turbulent transition to become an interesting field of investigation for the non-equilibrium statistical physics community.

\paragraph{Acknowledgements.}
This work was performed in the framework of the ``Arts and Sciences'' interdisciplinary project supported by the IXXI Complex System Institute (\emph{Institut Rh\^one-Alpin des Syst\`emes Complexes}) in Lyon, France.

\appendix

\section{Qualitative analogy with the Navier-Stokes equation}
\label{sec:app:A}

The form Eq.~(\ref{eq:transfer:DA}) of the energy transfer term, obtained by dimensional analysis, is reminiscent of the transfer terms in the Navier-Stokes equation.
Defining $\hat{\mathbf{v}}_\mathbf{k}$ the spatial Fourier transform of the velocity field $v(\mathbf{r})$, and the (three-dimensional) energy spectrum $\mE_\mathbf{k}=\frac{1}{2} |\hat{\mathbf{v}}_\mathbf{k}|^2$,
one has from the Navier-Stokes equation
\begin{equation}
    \frac{d\mE_{\mathbf{k}}}{dt} = - \nu k^2 \mE_{\mathbf{k}} - \sum_{\mathbf{k}'} \big(i (\mathbf{k}-\mathbf{k}')\cdot \hat{\mathbf{v}}_{\mathbf{k}'}\big)\, (\hat{\mathbf{v}}_\mathbf{k} \cdot \hat{\mathbf{v}}_{\mathbf{k}-\mathbf{k}'}) + i\mathbf{k} \frac{\hat{p}_{\mathbf{k}}}{\rho},
\end{equation}
with $k=|\mathbf{k}|$, and where $\hat{p}_{\mathbf{k}}$ is the spatial Fourier transform of the pressure (which ensures incompressibility), and $\rho$ is the constant density of the fluid \cite{Frisch}.

The qualitative analogy with our model goes as follows. The shell energy $E_i$ corresponds to the sum of $\mE_{\mathbf{k}}$ over all wavevectors $\mathbf{k}$ within shell $i$.
By splitting the sum over $\mathbf{k}'$ into a sum over $\mathbf{k}'$ with $k'<k$ and a sum over $\mathbf{k}'$ with $k'>k$, we obtain transfer terms from lower to higher wavenumbers.
These transfer terms depend on the velocity field $v_{\mathbf{k}'}$ and not only on the energy $\mE_{\mathbf{k}'}$. They thus contain information about both the vectorial nature of the flow
and the complex phase of the Fourier transform. However, at a heuristic level, we recover that the transfer term is proportional to the wavenumber $||\mathbf{k}-\mathbf{k}'||$ (instead of $k$ in our model), and that the dimension of the term 
$\hat{\mathbf{v}}_{\mathbf{k}'}\, (\hat{\mathbf{v}}_\mathbf{k} \cdot \hat{\mathbf{v}}_{\mathbf{k}-\mathbf{k}'})$ is the same as $\mE_{\mathbf{k}}^{3/2}$.
The function $f(x)$ in our model thus effectively encodes the average `interference' effects resulting from the complex phase of $v_{\mathbf{k}}$, and its vectorial nature.

\section{Harmonic synthesizer for the relaxation of the turbulent state}
\label{sec:app:B}

A harmonic synthesizer was built using FAUST \cite{faust} to generate dynamic sounds following the evolution of the solutions of Eq.~(\ref{eq:dEi:dt:full}). The equation was solved using a finite element method for $N=11$ energy modes, keeping $E_{11}$ fixed to the value $E_{11}=0.1$.
Parameter values are given in Table~\ref{table:appB}.
The synthesizer was designed as a sine wave of frequency $440$Hz and its $9$ first harmonics, where the energy modes $E_1$ to $E_{10}$ modulate the amplitude of each frequency
according to:
\begin{equation}
    \mathrm{synthesis}(t) = \sum_{i=1}^{10} \sin\left(\frac{440i}{2\pi}t\right) E_i(t).
\end{equation}
At $t=0$, energy was introduced in the system by setting each energy mode to $E_i=0.1$, and then the evolution under the stochastic dynamics Eq.~(\ref{eq:dEi:dt:full})
was recorded. We report in the movie (see Ancillary Files) the synthesized signal with different Reynolds numbers ($\Rey=0.93$, $1.85$, $3.09$, $30.9$) below and above the transition to turbulence,
and with different noise amplitudes ($\eta = 1$ or $2$). 
The video shows the spectrum over time of the different syntheses while they play.

\begin{table}[h!] \label{table:appB}
\caption{Parameter values used in the harmonic synthesizer.}
\begin{tabular}{|c|c|c|c|c|c|}
\hline
$a$ & $b$ & $\alpha$ & $\beta$ & $\lambda$ & $\nu$  \\
\hline 
1 & 1 & 0.1 & 0.93 & 1.08 & 1 \\
\hline
\end{tabular}
\end{table}


\bibliographystyle{plain}


\end{document}